\documentclass{aa}

\usepackage{graphicx}
\usepackage{epsfig}
\usepackage{natbib}
\usepackage{txfonts} 
\usepackage{hyperref}
\usepackage{supertabular}

\newcommand{\COBOLD}{{\tt CO$^5$BOLD}}

\newcommand{\LHD}{{\tt LHD}}
\newcommand{\MARCS}{{\tt MARCS}}
\newcommand{\ATLAS}{{\tt ATLAS9}}
\newcommand{\Linfor}{{\tt Linfor3D}}
\newcommand{\NLTE}{{\tt NLTE3D}}

\newcommand{\moh}{\ensuremath{[\mathrm{M/H}]}}
\newcommand{\aoFe}{\ensuremath{\left[\mathrm{\alpha}/\mathrm{Fe}\right]}}\newcommand{\teff}{\ensuremath{T_{\mathrm{eff}}}}

\newcommand{\xtmean}[1]{\ensuremath{\left\langle #1\right\rangle}}

\begin{document}

\title{Lithium spectral line formation in stellar atmospheres}
   \subtitle{The impact of convection and NLTE effects}

\author{
        J. Klevas\inst{1}
        \and
        A. \,Ku\v{c}inskas\inst{1}
        \and
        M. \,Steffen\inst{2,3}
        \and
        E. \,Caffau\inst{3}
        \and
        H.-G. \,Ludwig\inst{4}
       }

\institute{
        Institute of Theoretical Physics and Astronomy, Vilnius University, A. Go\v {s}tauto 12, Vilnius LT-01108, Lithuania \\
        \email{jonas.klevas@tfai.vu.lt}
        \and
        Leibniz-Institut f\"ur Astrophysik Potsdam, An der Sternwarte 16, D-14482 Potsdam, Germany
        \and
        GEPI, Observatoire de Paris, CNRS, Universit\'{e} Paris Diderot, Place Jules Janssen, 92190 Meudon, France
        \and
        ZAH Landessternwarte K\"{o}nigstuhl, D-69117 Heidelberg, Germany
}


\abstract
{}
{Because of the complexities involved in treating spectral line formation in full 3D and non-local thermodynamic equilibrium (NLTE), different simplified approaches are sometimes used to account for the NLTE effects with 3D hydrodynamical model atmospheres. In certain cases, chemical abundances are derived in 1D\,NLTE and then corrected for the 3D effects by adding 3D--1D\,LTE (Local Thermodynamic Equilibrium, LTE) abundance corrections (3D+NLTE approach). Alternatively, average $\xtmean{\mbox{3D}}$ model atmospheres are sometimes used to substitute for the full 3D hydrodynamical models.}
{In this work we tested whether the results obtained using these simplified schemes (3D+NLTE, $\xtmean{\mbox{3D}}$\,NLTE) may reproduce those derived using the full 3D\,NLTE computations. The tests were made using 3D hydrodynamical \COBOLD\ model atmospheres of the main sequence (MS), main sequence turn-off (TO), subgiant (SGB), and red giant branch (RGB) stars, all at two metallicities, $\moh=0.0$ and $-2.0$. Our goal was to investigate the role of 3D and NLTE effects on the formation of the 670.8~nm lithium resonance line. This was done by assessing differences in the strengths of synthetic 670.8~nm line profiles, which were computed using 3D/1D NLTE/LTE approaches.}
{Our results show that Li 670.8~nm line strengths obtained using different methodologies differ only slightly in most of the models at solar metallicity studied here. However, the line strengths predicted with the 3D\,NLTE and 3D+NLTE approaches  become significantly different at subsolar metallicities. At $\moh=-2.0$, this may lead to (3D\,NLTE)\,--\,(3D+NLTE) differences in the predicted lithium abundance of $\sim0.46$ and $\sim0.31$~dex in the TO and RGB stars respectively. On the other hand, NLTE line strengths computed with the average $\xtmean{\mbox{3D}}$ and 1D model atmospheres are similar to those obtained with the full 3D\,NLTE approach for MS, TO, SGB, and RGB stars, at all metallicities; ${\rm 3D}-\xtmean{\mbox{3D}}$ and ${\rm 3D}-{\rm 1D}$ differences in the predicted abundances are always less than $\sim0.04$~dex and $\sim0.08$~dex, respectively. However, neither of the simplified approaches can reliably substitute 3D\,NLTE spectral synthesis when precision is required.}
{}


\authorrunning{Klevas et al.}
\titlerunning{Convection and NLTE effects in lithium line formation within stellar atmospheres, accuracy of alternate approaches}

\maketitle

\section{Introduction}

Spectral lines in stellar atmospheres frequently form in conditions that may deviate significantly from local thermodynamic equilibrium (LTE). This is normally taken into account by solving the statistical equilibrium equations, coupled with non-LTE (NLTE) radiative transfer computations. Fortunately, except for the hottest stars, NLTE effects have relatively minor influence on the thermodynamic structures of the model atmospheres \citep[e.g.,][]{SH06}. However, in many cases the NLTE treatment in the computation of synthetic spectral line profiles is very important, especially at lower metallicities where NLTE effects may be significant \citep[see, e.g.,][for a review]{A05}. So far, however, attempts to include NLTE effects have been mostly limited to using 1D hydrostatic model atmospheres.

At the same time, recent studies made with the 3D hydrodynamical model atmospheres in LTE have revealed that various dynamical phenomena may also be  very important in the spectral line formation taking place in convective stellar atmospheres \citep[e.g.,][]{ANTS99,CAT07,CAN09,BBL10,BCR13,DKS13,MCH13}. It would therefore be  desirable if abundances of chemical elements in stellar atmospheres could be derived using the full 3D\,NLTE approach, i.e., with NLTE spectral synthesis computations performed using 3D hydrodynamical model atmospheres, which would be the most realistic way to model spectral line formation. This, however, has rarely been possible  since the majority of current spectral synthesis codes lack the capability to perform the statistical equilibrium computations in 3D\,NLTE, even though several successful steps in this direction have already been made, with very likely more to come in the near future (see, for example, \citealt{ACB03,AGS04,CSC07,SCB10,SCC12,LAC12,LMA13,HS13,PSK13,SPC15, AAC15}).  Frequently then, various simplifications are still used to substitute for the full 3D NLTE approach. For example, sometimes elemental abundances are derived in 1D\,NLTE and subsequently corrected for the 3D hydrodynamical effects, with the size of these corrections evaluated under the assumption of LTE (3D+NLTE approach); or where average $\xtmean{\mbox{3D}}$ model atmospheres are used with the standard 1D\,NLTE analysis tools. However, since the properties of spectral line formation in full 3D\,NLTE are still poorly explored, many details of the interaction between hydrodynamical and NLTE effects in 3D are still unknown. Therefore, it is not clear whether such strategies may be generalized to substitute for the full 3D\,NLTE computations.

Lithium is an important cosmological tracer element, and one way of measuring its primordial (cosmological) abundance is from stellar spectra (mostly of MS, TO, and SGB stars; however, RGB stars, may be useful too: see, for example, \citealt{MSB12}). Lithium is known to be prone to both 3D hydrodynamical and NLTE effects, and it is thought that these effects typically tend to work in opposite directions in stellar atmospheres, at least in those of MS to SGB stars, which leads to generally small 3D\,NLTE abundance corrections \citep[][]{ACB03,SCB10}. Nevertheless, it is still not entirely clear  whether less sophisticated tools/approaches could be used in the case of lithium to circumvent the full 3D\,NLTE calculations, which are still computationally expensive and less accessible to wider astronomical community.

In this paper, we therefore investigate whether the 3D+NLTE and/or $\xtmean{\mbox{3D}}$\,NLTE approaches could be used to substitute the full 3D\,NLTE methodology to obtain 3D\,NLTE abundances of lithium in the atmospheres of MS, TO, SGB, and RGB stars.

The tests are performed using state-of-the-art 3D hydrodynamical \COBOLD\ and 1D hydrostatic \LHD\ model atmospheres, computed at two different metallicities, $\moh = 0.0$ and $-2.0$. The \COBOLD\ and \LHD\ model atmospheres share identical atmospheric parameters, chemical composition, equation of state, and opacities, to make a strictly differential analysis of the role of 3D/1D and NLTE/LTE effects in the spectral line formation.

The paper is structured as follows: In Sect.~\ref{sect:method} we describe the model atmospheres utilized in our study and outline the methodology of 3D/1D\,NLTE/LTE spectral synthesis computations. The results obtained using various spectral line synthesis approaches are presented and discussed in Sect.~\ref{sect:results}, while in Sect.~\ref{sect:conclus} we provide a short summary of the most important results obtained in this work.

\section{Methodology\label{sect:method}}

\subsection{Model atom of lithium\label{sect:modatom}}

\begin{figure}[tb]
\centering
\mbox{\includegraphics[width=8.5cm]{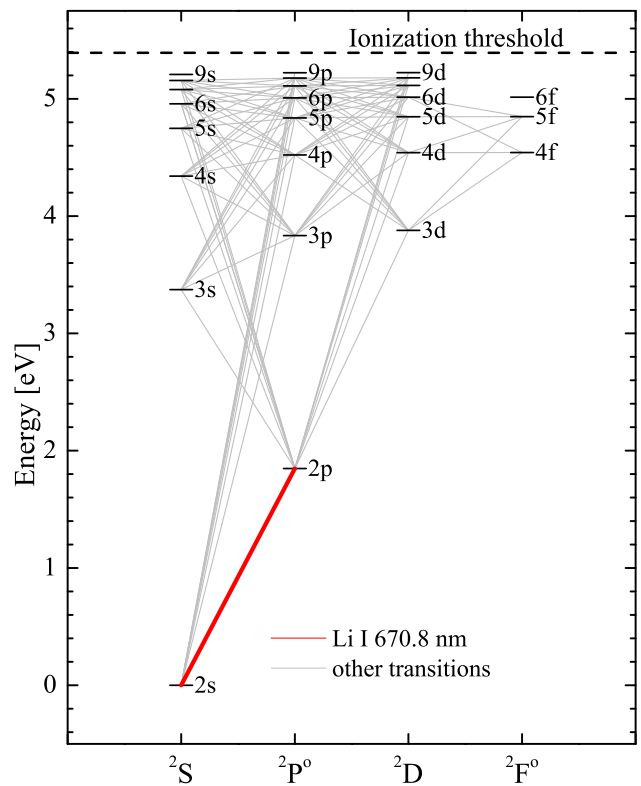}}
\caption{
Model atom of lithium used in the 3D\,NLTE spectral line synthesis computations. The thick red line indicates the transition that corresponds to the lithium 670.8 nm resonance doublet. Other radiative bound-bound transitions are shown as grey lines.}
\label{fig:grotrian}
\end{figure}

In this study, we use a model atom of \ion{Li}{i} which was originally developed and tested by \citet{CSC07} and \citet{SBC10}. For the purposes of the current work, the model atom was updated and now consists of 26 levels and 123 (96 of which are radiative) bound-bound transitions of \ion{Li}{i} and the ground level of \ion{Li}{ii}, with each level of \ion{Li}{i} coupled to the continuum via bound-free transitions. (The ground state of \ion{Li}{ii} in the current model atom is always in LTE, since lithium is mostly fully ionized throughout the model atmospheres studied in this work.) This renders the model atom complete up to the principal quantum number $n=6$ and spectroscopic term $^2{\rm F}^{\rm o}$, with additional energy levels up to $n=9$ and term $^2{\rm D}$ (Fig.~\ref{fig:grotrian}). Data concerning atomic energy levels and transitions (level energies and statistical weights; wavelengths and Einstein coefficients of the bound-bound transitions) were taken from the NIST database. We used electron collisional excitation and ionization rates from the quantum mechanical computations of \citet[][]{OBL11} for the energy levels of up to 5s ($^2{\rm S}$). Elsewhere, collisional excitation by electrons for radiatively permitted transitions was accounted for by using the classical formula of \citet[][]{VR62}, while the formula of \citet[][]{S62} was used to compute collisional electron ionization rates. To account for the collisional excitation by hydrogen, we used collisional excitation rates computed by \citet[][]{BBA03}, while the classical formula of Drawin (in the formulation of \citealt[][]{L93}) was used for radiatively permitted transitions when no quantum mechanical data were available. Hydrogen H--Li charge transfer rates were taken from \citet[][]{BBA03} for the atomic levels up to 4p inclusive. Bound-free transitions resulting from collisions with hydrogen were expected to be inefficient and thus were ignored. Photoionization cross sections were taken from TOPBASE \citep[][]{CMO93}. No scaling of collisional rates was applied in the calculations of bound-free and bound-bound transitions. Information about the energy levels and bound-bound radiative transitions, included in the present version of the \ion{Li}{i} model atom, are provided in Appendix~\ref{app:atom}, Tables~\ref{tab:Liatom_levels}, and \ref{tab:Liatom_transitions}, respectively. Twenty-seven transitions in the model atom are purely collisional. Collisional radiatively-forbidden transitions involving \ion{Li}{i} levels beyond 5s were not accounted for since reliable quantum-mechanical data for these transitions are not available. We note that the role of the omitted transitions between the higher levels is minor: when they are taken into account using the formula of \citet[]{A73}, collision strength $\Omega = 1$, the change in the estimated abundance (which directly applies to abundance corrections, too) is always less than $0.05$~dex, with typical values being significantly smaller.

\subsection{3D and 1D model atmospheres, spectral line synthesis, and abundance corrections\label{sect:specsynth}}

The 3D hydrodynamical models used in this work were taken from the CIFIST \COBOLD\ model atmosphere grid \citep{LCS09}. All simulation runs cover $\sim13-480$ convective turnover times, as measured by the Brunt-Vais\"{a}l\"{a} timescale \citep[see][for the definition]{KSL13}. The models were computed using solar-scaled elemental abundances from \citet{AGS05}, with a constant enhancement of $\aoFe=+0.4$ in alpha-element abundances for models at $\moh=-2.0$. Monochromatic opacities were taken from the \MARCS\ model atmosphere package \citep{GEK08} and were grouped into five and six opacity bins \citep{LCS09} for models at $\moh=0.0$ and $-2.0$, respectively. All simulations were performed under the assumption of LTE using a cartesian model grid; the gravity vector has been assumed constant and antiparallel to $z$-axis throughout the model box, thus the effects of sphericity were ignored \citep[for more details on the model calculations see][]{LCS09}. For comparison, we also used 1D hydrostatic \LHD\ model atmospheres \citep[][]{CLS08}. It is important to note that both \COBOLD\ and \LHD\ models shared identical atmospheric parameters, opacities, chemical composition, and equation of state, which allowed us to make a strictly differential comparison of their predictions.

\begin{figure}[tb]
\centering
\includegraphics[width=9.cm]{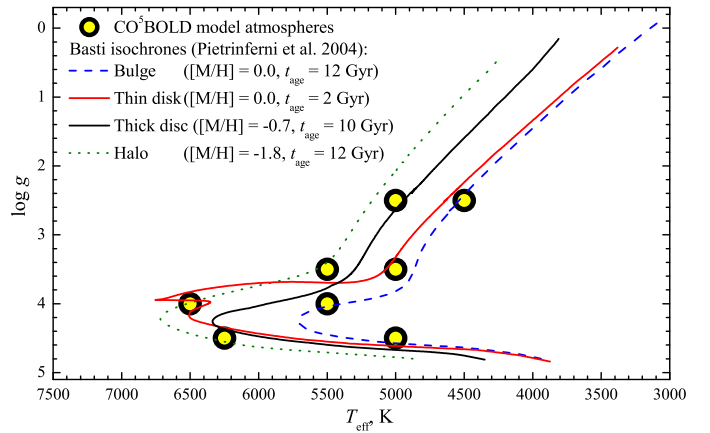}
\caption{
3D hydrodynamical \COBOLD\ model atmospheres used in this study, plotted in the $\log g - \teff$ plane (large filled circles). Several isochrones are shown as lines to indicate the approximate loci of stars in various Galactic populations.
}
\label{fig:isochrones}
\end{figure}

We used 3D hydrodynamical and 1D hydrostatic models of the representative RGB, SGB, TO, and MS stars, in each case computed for the same $\log g$ at two different effective temperatures and metallicities (see Table~\ref{tab:3Dmodels}). This choice of model parameters allowed us to bracket the range of \teff, $\log g$, and \moh\ typical of real stars observed in various Galactic populations. At the same time, this provides an opportunity to obtain a rough estimate of the size of 3D/1D and/or NLTE/LTE effects in real stellar atmospheres that are covered by our parameter range, by interpolating our results obtained at the bracketing values of stellar parameters. Atmospheric parameters of the 3D hydrodynamical \COBOLD\ model atmospheres used in our study are provided in Table~\ref{tab:3Dmodels}, with their positions in the $\log g - \teff$ plane shown in Fig.~\ref{fig:isochrones}. At each point in the $\log g - \teff$ diagram, calculations were performed with two models of different metallicity, $\moh =0.0$ and $-2.0$, to assess the differential effect of metallicity on the lithium spectral line formation.

\begin{table}[tb]
\caption{3D hydrodynamical \COBOLD\ model atmospheres used in this work.\label{tab:3Dmodels}}
\centering
\setlength{\tabcolsep}{2pt}
\begin{tabular}{lccccc}
\hline
 Model   & \teff & $\log g$ &    \moh\    &      Grid size , Mm       &      Grid resolution      \\
         &   K   &          &             &   $x \times y \times z$   &    $x\times y\times z$    \\
\hline
 RGB \#1 & 4480  & 2.5      & $\;\;\,0.0$ & 851$\times$851$\times$295 & 140$\times$140$\times$150 \\
 RGB \#2 & 4970  & 2.5      & $\;\;\,0.0$ & 573$\times$573$\times$243 & 160$\times$160$\times$200 \\
 RGB \#3 & 4480  & 2.5      & $-2.0$      & 851$\times$851$\times$292 & 140$\times$140$\times$150 \\
 RGB \#4 & 5020  & 2.5      & $-2.0$      & 584$\times$584$\times$245 & 160$\times$160$\times$200 \\
 SGB \#1 & 4920  & 3.5      & $\;\;\,0.0$ & 59.7$\times$59.7$\times$30.2 & 140$\times$140$\times$150 \\
 SGB \#2 & 5430  & 3.5      & $\;\;\,0.0$ & 49.0$\times$49.0$\times$35.9 & 140$\times$140$\times$150 \\
 SGB \#3 & 4980  & 3.5      & $-2.0$      & 59.7$\times$59.7$\times$30.2 & 140$\times$140$\times$150 \\
 SGB \#4 & 5500  & 3.5      & $-2.0$      & 49.0$\times$49.0$\times$35.9 & 140$\times$140$\times$150 \\
 TO \#1  & 5480  & 4.0      & $\;\;\,0.0$ & 20.3$\times$20.3$\times$10.6    & 140$\times$140$\times$150 \\
 TO \#2  & 6490  & 4.0      & $\;\;\,0.0$ & 29.0$\times$29.0$\times$14.9    & 140$\times$140$\times$150 \\
 TO \#3  & 5470  & 4.0      & $-2.0$      & 20.1$\times$20.1$\times$10.6    & 140$\times$140$\times$150 \\
 TO \#4  & 6530  & 4.0      & $-2.0$      & 29.6$\times$29.6$\times$14.9    & 140$\times$140$\times$150 \\
 MS \#1  & 6230  & 4.5      & $\;\;\,0.0$ & 7.00$\times$7.00$\times$4.02 & 140$\times$140$\times$150 \\
 MS \#2  & 4980  & 4.5      & $\;\;\,0.0$ & 4.94$\times$4.94$\times$2.48 & 140$\times$140$\times$141 \\
 MS \#3  & 6320  & 4.5      & $-2.0$      & 7.00$\times$7.00$\times$4.02 & 140$\times$140$\times$150 \\
 MS \#4  & 5010  & 4.5      & $-2.0$      & 5.08$\times$5.08$\times$2.49 & 140$\times$140$\times$141 \\
\hline
\end{tabular}
\end{table}

An updated version of the \NLTE\ code\footnote{For details on the \NLTE\ code see \citet{SBC10,PSK13}.} was used to compute the NLTE departure coefficients, $b_{\rm i}(x,y,z)$, at each geometrical position $(x,y,z)$ in the background model atmosphere for each energy level $i$, utilizing the model atom of \ion{Li}{i} described in the previous section ($b_{\rm i}(x,y,z)=n_{\rm i}(x,y,z)^{\rm NLTE}/n_{\rm i}(x,y,z)^{\rm LTE}$, where $n_{\rm i}(x,y,z)^{\rm NLTE}$ and $n_{\rm i}(x,y,z)^{\rm NLTE}$ are population densities of a given level $i$ in NLTE and LTE, respectively). NLTE3D computes the departure coefficients by solving the statistical equilibrium equations at each grid cell $(x,y,z)$ using an accelerated $\Lambda$-iteration scheme. The line-blanketed mean continuum intensity, $J_{\nu}(x,y,z)$, is computed using the {\tt BIG} opacity distribution functions (ODFs) from the \ATLAS\ model atmosphere package, at 600 frequency points, which spans the spectral range from 92.5 to 7300~nm, and accounts for coherent, isotropic scattering in the continuum. $J_\nu$ was computed as a weighted average of the intensities obtained along the vertical direction, four azimuthal directions, and four inclination angles (i.e., 17 rays in total).

The departure coefficients were computed using three types of model atmospheres: (i) 3D hydrodynamical, calculated with the \COBOLD\ code \citep[][]{FSL12}; (ii) $\xtmean{\mbox{3D}}$ average atmospheres, obtained by taking a temporal and spatial average over the sequence of full 3D model structures; and (iii) 1D hydrostatic atmospheres, calculated using the \LHD\ code \citep[][]{CLS08}. To calculate $\xtmean{\mbox{3D}}$ average atmospheres, we averaged the fourth moment of temperature and the first moment of gas pressure over surfaces of constant Rosseland optical depth, according to the prescription provided in \citet[][]{SLF95}. For detailed reasoning behind the choice of the averaging scheme, see, e.g., Appendix A of \citet{DKS13}. While this choice is fully adequate for the purposes of the current study, there are other options for constructing average $\xtmean{\mbox{3D}}$ model atmospheres, depending on the specific task \citep[see, e.g.,][]{MCH13}. The average $\xtmean{\mbox{3D}}$ model is in fact a 1D model and, because it is obtained by temporal and horizontal averaging of the full 3D model sequence, it lacks information about the horizontal inhomogeneities that are present in the 3D model structures. Thus, the comparison of the line formation properties in full 3D and $\xtmean{\mbox{3D}}$ models allows us to assess the importance of horizontal inhomogeneities in the process of line formation.

In most astrophysical situations, the only spectral line available for lithium diagnostics in the UV--IR stellar spectra is the \ion{Li}{i} 670.8\,nm resonance doublet. We therefore focused on this single tracer by computing its synthetic 3D and 1D profiles with the \Linfor\ spectral synthesis code, both in LTE and NLTE. The doublet structure of the \ion{Li}{i} 670.8 nm resonance line was accounted for when computing the bound-bound radiative rates, while assuming identical NLTE departure coefficients for the two fine structure sublevels of the 2p level. To synthesize the line profile, the fine structure of the \ion{Li}{i} 670.8 nm feature  was taken into account.  The analysis was done in the cases of (i) weak (equivalent width $W=0.5$~pm), and (ii) strong ($W=8$~pm) lines. These line strengths bracket the typical EWs of the \ion{Li}{i} 670.8\,nm line observed in stars with atmospheric parameters similar to those studied in this work. However, as we will see in Sect.~\ref{sect:results}, there is little sensitivity of the abundance corrections to the line strength for all model atmospheres studied here.

The interplay between 3D hydrodynamical and NLTE effects was studied with the help of abundance corrections \citep[see, e.g.,][]{CLS11}. The latter were defined as differences in Li abundance that would be determined from the \ion{Li}{i} 670.8\,nm line of a given strength using different model atmospheres, in NLTE and/or LTE. First, we computed the following abundance corrections: $\Delta_{{\rm 3D\,NLTE}~-~{\rm 1D\,LTE}}$, $\Delta_{{\rm 1D\,NLTE}~-~{\rm 1D\,LTE}}$, and $\Delta_{{\rm 3D\,LTE}~-~{\rm 1D\,LTE}}$. Then, the latter two corrections were used to obtain the $\Delta_{\rm 3D~+\,NLTE} \equiv \Delta_{{\rm 1D\,NLTE}~-~{\rm 1D\,LTE}} + \Delta_{{\rm 3D\,LTE}~-~{\rm 1D\,LTE}}$ abundance correction, i.e., the total correction expected when the 1D\,NLTE abundances are corrected for 3D effects by adding the 3D--1D\,LTE abundance correction. Finally, we also calculated the $\Delta_{{\rm \langle3D\rangle\,NLTE - 1D\,LTE}}$ abundance correction obtained by using the average $\xtmean{\mbox{3D}}$ and 1D model atmospheres. All obtained abundance corrections are listed in Table~\ref{tab:abucor}.

Finally, we note that in this work we frequently used 1D\,LTE abundance estimates as the reference point for computing and assessing various abundance corrections. This choice may be well justified if one is interested in knowing how much the lithium abundances, which were obtained using various more sophisticated approaches, would differ from those determined in 1D\,LTE. At the same time, this may offer a convenient way to ``correct'' the 1D\,LTE abundances for 3D and/or NLTE effects. Nevertheless, it is clearly the full 3D\,NLTE approach that provides the highest realism in modeling lithium spectral line formation in stellar atmospheres. As such, ideally the 3D\,NLTE methodology should be applied to obtain the most reliable lithium abundances from the measured equivalent widths of the \ion{Li}{i} 670.8 nm lines (e.g., by using the 3D NLTE - 1D LTE abundance corrections).

\section{Results and discussion\label{sect:results}}

\begin{figure}[!tb]
\centering
\includegraphics[width=8.50cm]{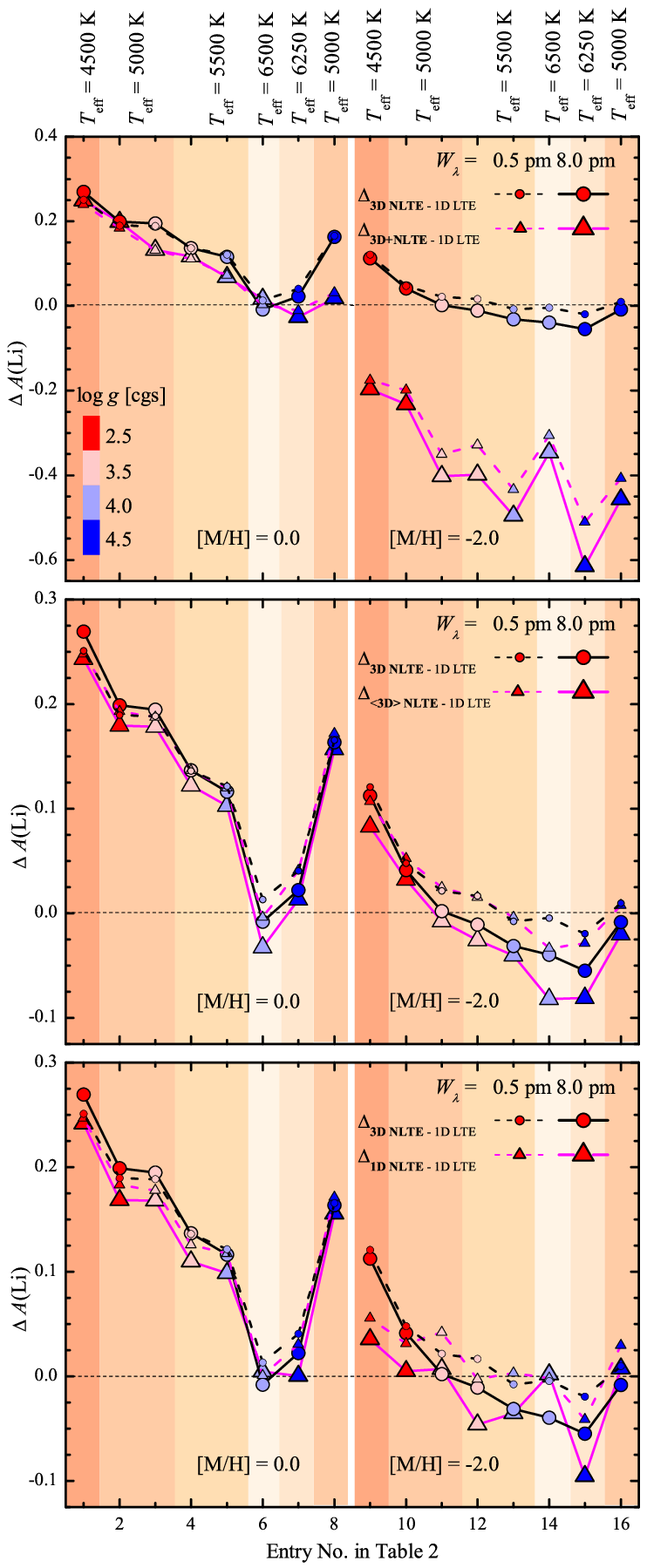}
\caption{
$\Delta_{{\rm 3D\,NLTE}~-~{\rm 1D\,LTE}}$ (shown in all panels), $\Delta_{{\rm 3D+NLTE}~-~{\rm 1D\,LTE}}$ (top), $\Delta_{{\rm \langle3D\rangle\,NLTE - 1D\,LTE}}$ (middle) and $\Delta_{{\rm 1D\,NLTE}~-~{\rm 1D\,LTE}}$  (bottom) abundance corrections for the lithium 670.8 nm line in the atmospheres of RGB, SGB, TO, and MS stars, at $\moh = 0.0$ and $-2.0$, in the order they appear in Table \ref{tab:abucor}.
}
\label{fig:abucor}
\end{figure}

\begin{table*}[!th]
\caption{Abundance corrections for the lithium 670.8 nm line in the
atmospheres of RGB, SGB, TO, and MS stars, at $\moh = 0.0$ and $-2.0$. The
microturbulence velocity used in the $\xtmean{\mbox{3D}}$ and 1D line
synthesis was set to 2~km/s. \label{tab:abucor}}
\centering
\setlength{\tabcolsep}{7pt}
\begin{tabular}{ccccccc}
\hline\noalign{\smallskip}
  Entry No. & $W$ & $\Delta_{{\rm 1D\,NLTE - 1D\,LTE}}$  & $\Delta_{{\rm 3D\,LTE - 1D\,LTE}}$ & $\Delta_{{\rm 3D + NLTE}}$
     & $\Delta_{{\rm 3D\,NLTE - 1D\,LTE}}$  & $\Delta_{{\rm \langle3D\rangle\,NLTE - 1D\,LTE}}$    \\
  & pm &             dex                &              dex            &               dex
     &             dex                &              dex               \\ (1) & (2) & (3) & (4) & (5) & (6) & (7) \\
\noalign{\smallskip}
1 & \multicolumn{5}{l}{RGB \#1: $\teff=4480$~K, $\log g=2.5$, $\moh=0.0$} & \\
&       0.5     &       0.247   &       -0.008  &       0.239   &       0.251   &       0.248   \\
&       8.0     &       0.242   &       0.007   &       0.249   &       0.269   &       0.243   \\
\noalign{\smallskip}
2 & \multicolumn{5}{l}{RGB \#2: $\teff=4970$~K, $\log g=2.5$, $\moh=0.0$} & \\
&       0.5     &       0.183   &       0.001   &       0.184   &       0.190   &       0.193   \\
&       8.0     &       0.168   &       0.030   &       0.199   &       0.199   &       0.180   \\
\noalign{\smallskip}
3 & \multicolumn{5}{l}{SGB \#1: $\teff=4920$~K, $\log g=3.5$, $\moh=0.0$} & \\
&       0.5     &       0.178   &       -0.047  &       0.131   &       0.189   &       0.187   \\
&       8.0     &       0.168   &       -0.036  &       0.132   &       0.195   &       0.178   \\
\noalign{\smallskip}
4 & \multicolumn{5}{l}{SGB \#2: $\teff=5430$~K, $\log g=3.5$, $\moh=0.0$} & \\
&       0.5     &       0.126   &       -0.015  &       0.111   &       0.136   &       0.136   \\
&       8.0     &       0.110   &       0.007   &       0.117   &       0.137   &       0.122   \\
\noalign{\smallskip}
5 & \multicolumn{5}{l}{TO \#1: $\teff=5480$~K, $\log g=4.0$, $\moh=0.0$} & \\
&       0.5     &       0.117   &       -0.045  &       0.072   &       0.121   &       0.120   \\
&       8.0     &       0.099   &       -0.030  &       0.068   &       0.116   &       0.103   \\
\noalign{\smallskip}
6 & \multicolumn{5}{l}{TO \#2: $\teff=6490$~K, $\log g=4.0$, $\moh=0.0$} & \\
&       0.5     &       0.001   &       0.002   &       0.003   &       0.013   &       -0.004  \\
&       8.0     &       0.004   &       0.012   &       0.016   &       -0.008  &       -0.033  \\
\noalign{\smallskip}
7 & \multicolumn{5}{l}{MS \#1: $\teff=6230$~K, $\log g=4.5$, $\moh=0.0$} & \\
&       0.5     &       0.030   &       -0.041  &       -0.011  &       0.040   &       0.041   \\
&       8.0     &       0.000   &       -0.026  &       -0.026  &       0.022   &       0.013   \\
\noalign{\smallskip}
8 & \multicolumn{5}{l}{MS \#2: $\teff=4980$~K, $\log g=4.5$, $\moh=0.0$} & \\
&       0.5     &       0.171   &       -0.140  &       0.031   &       0.166   &       0.172   \\
&       8.0     &       0.155   &       -0.137  &       0.019   &       0.163   &       0.157   \\
\noalign{\smallskip}
9 & \multicolumn{5}{l}{RGB \#3: $\teff=4480$~K, $\log g=2.5$, $\moh=-2.0$} & \\
&       0.5     &       0.056   &       -0.231  &       -0.176  &       0.121   &       0.107   \\
&       8.0     &       0.035   &       -0.232  &       -0.197  &       0.113   &       0.083   \\
\noalign{\smallskip}
10 & \multicolumn{5}{l}{RGB \#4: $\teff=5020$~K, $\log g=2.5$, $\moh=-2.0$} & \\
&       0.5     &       0.031   &       -0.230  &       -0.199  &       0.048   &       0.052   \\
&       8.0     &       0.005   &       -0.237  &       -0.232  &       0.041   &       0.032   \\
\noalign{\smallskip}
11 & \multicolumn{5}{l}{SGB \#3: $\teff=4980$~K, $\log g=3.5$, $\moh=-2.0$} & \\
&       0.5     &       0.042   &       -0.393  &       -0.351  &       0.021   &       0.025   \\
&       8.0     &       0.007   &       -0.409  &       -0.402  &       0.002   &       -0.008  \\
\noalign{\smallskip}
12 & \multicolumn{5}{l}{SGB \#4: $\teff=5500$~K, $\log g=3.5$, $\moh=-2.0$} & \\
&       0.5     &       -0.003  &       -0.325  &       -0.328  &       0.017   &       0.015   \\
&       8.0     &       -0.046  &       -0.352  &       -0.398  &       -0.011  &       -0.026  \\
\noalign{\smallskip}
13 & \multicolumn{5}{l}{TO \#3: $\teff=5470$~K, $\log g=4.0$, $\moh=-2.0$} & \\
&       0.5     &       0.003   &       -0.437  &       -0.434  &       -0.008  &       -0.004  \\
&       8.0     &       -0.036  &       -0.460  &       -0.495  &       -0.031  &       -0.040  \\
\noalign{\smallskip}
14 & \multicolumn{5}{l}{TO \#4: $\teff=6530$~K, $\log g=4.0$, $\moh=-2.0$} & \\
&       0.5     &       -0.002  &       -0.304  &       -0.306  &       -0.004  &       -0.004  \\
&       8.0     &       0.002   &       -0.348  &       -0.346  &       -0.040  &       -0.082  \\
\noalign{\smallskip}
15 & \multicolumn{5}{l}{MS \#3: $\teff=6320$~K, $\log g=4.5$, $\moh=-2.0$} & \\
&       0.5     &       -0.040  &       -0.470  &       -0.510  &       -0.019  &       -0.029  \\
&       8.0     &       -0.094  &       -0.520  &       -0.614  &       -0.055  &       -0.081  \\
\noalign{\smallskip}
16 & \multicolumn{5}{l}{MS \#4: $\teff=5010$~K, $\log g=4.5$, $\moh=-2.0$} & \\
&       0.5     &       0.030   &       -0.437  &       -0.407  &       0.010   &       0.008   \\
&       8.0     &       0.007   &       -0.464  &       -0.457  &       -0.008  &       -0.020  \\
\noalign{\smallskip}
\hline
\end{tabular}
\end{table*}

\begin{figure*}[tb]
\centering
\mbox{\includegraphics[width=7.5cm]{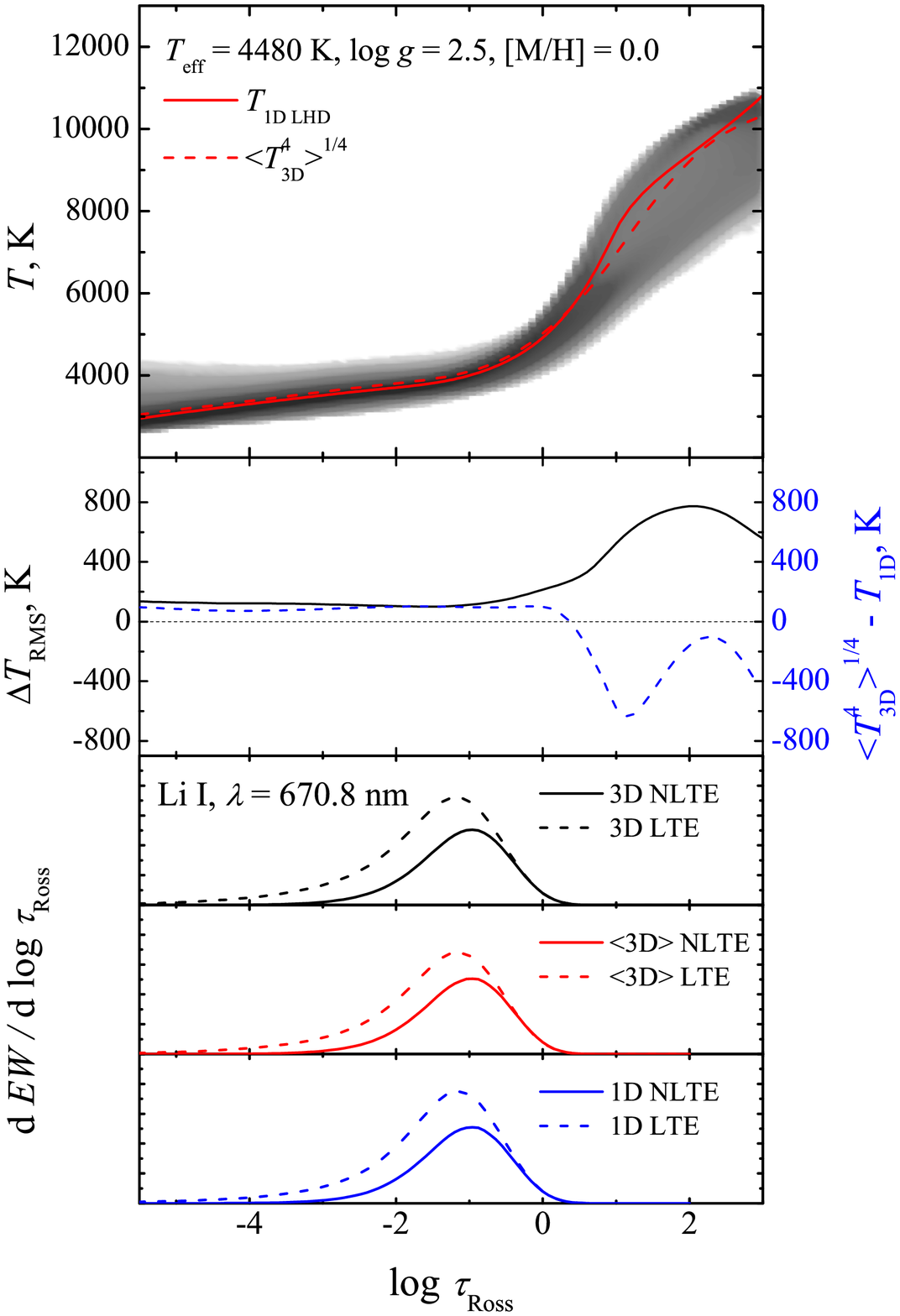}}
\mbox{\includegraphics[width=7.5cm]{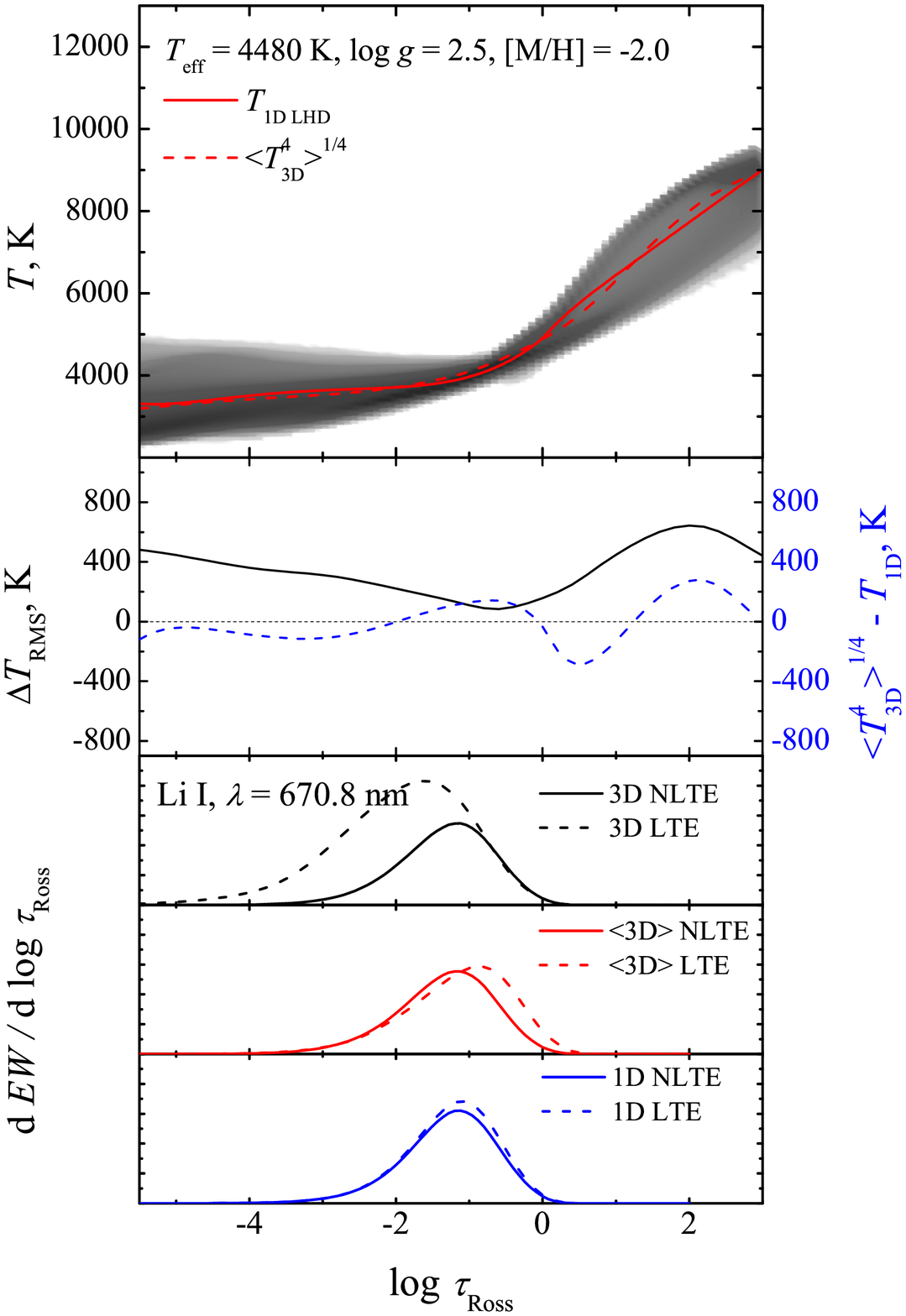}}
\caption{\footnotesize
Thermodynamic structures and Li 670.8 nm line formation properties in two model atmospheres of red giant stars ($\teff=4480$~K, $\log g=2.5$), at $\moh=0.0$ (left) and $\moh=-2.0$ (right). \textbf{From top to bottom: (1)} temperature profiles in the 3D hydrodynamical (probability density, grayscale), average $\xtmean{\mbox{3D}}$ (dashed red line), and 1D~LHD (solid red line) model atmospheres; \textbf{(2)} horizontal temperature fluctuations in the 3D model ($\Delta T_{\rm RMS}$, solid black line) and differences between the temperature profiles of the $\xtmean{\mbox{3D}}$ and 1D~LHD model atmospheres (blue dashed lines); \textbf{(3--5)} contribution functions of the Li 670.8 nm line (i.e., rates of the line equivalent width growth, $\mathrm{d}EW/\mathrm{d}\log{\tau_\mathrm{Ross}}$, as a function of optical depth) in 3D\,NLTE/LTE (black), $\xtmean{\mbox{3D}}$\,NLTE/LTE (red), and 1D\,NLTE/LTE (blue). For each metallicity, the lithium abundance is fixed such that $W({\rm 1D\,LTE}) = 0.5$~pm.}
\label{fig:struct-g25}
\end{figure*}

\begin{figure*}[tb]
\centering
\mbox{\includegraphics[width=7.5cm]{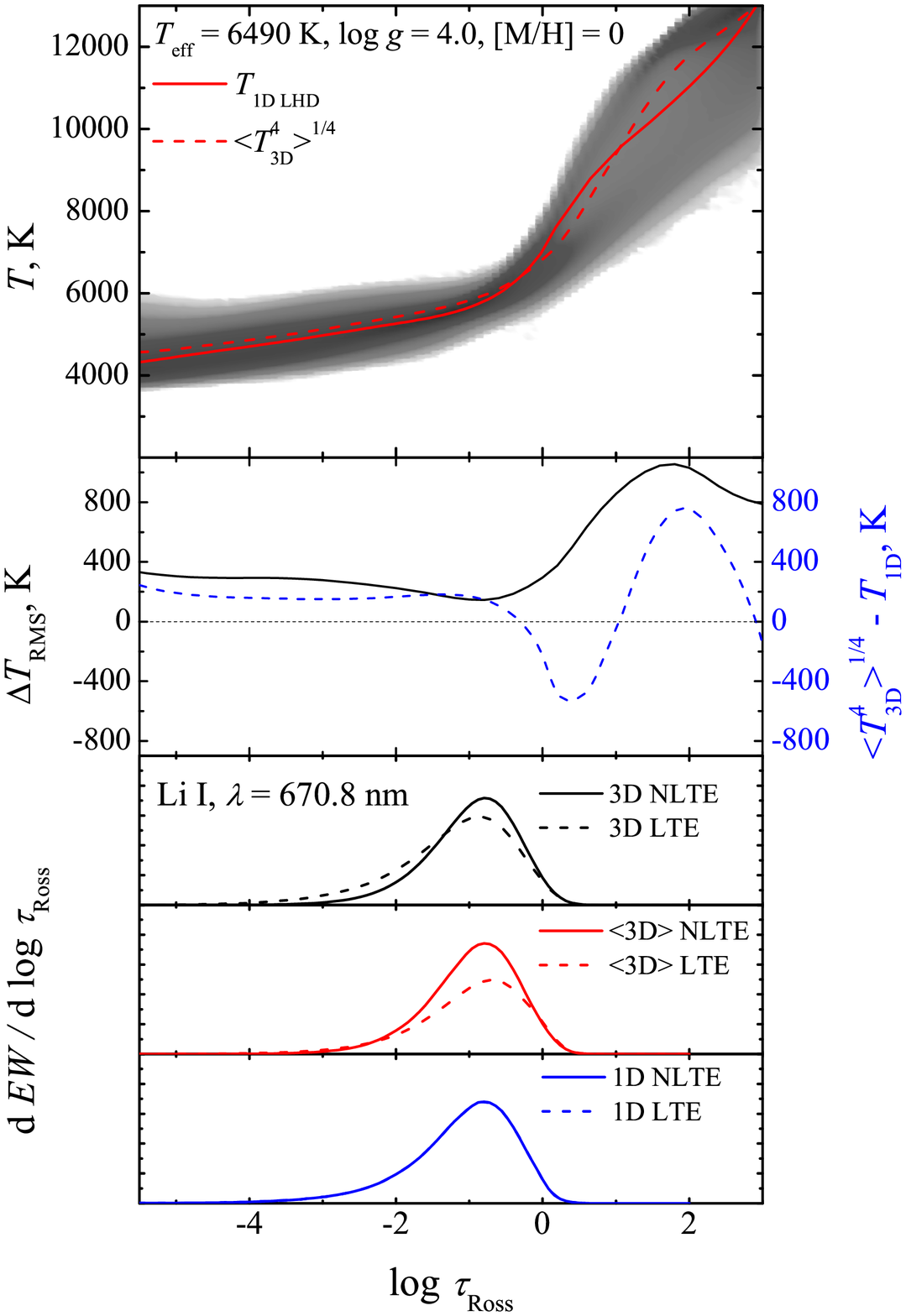}}
\mbox{\includegraphics[width=7.5cm]{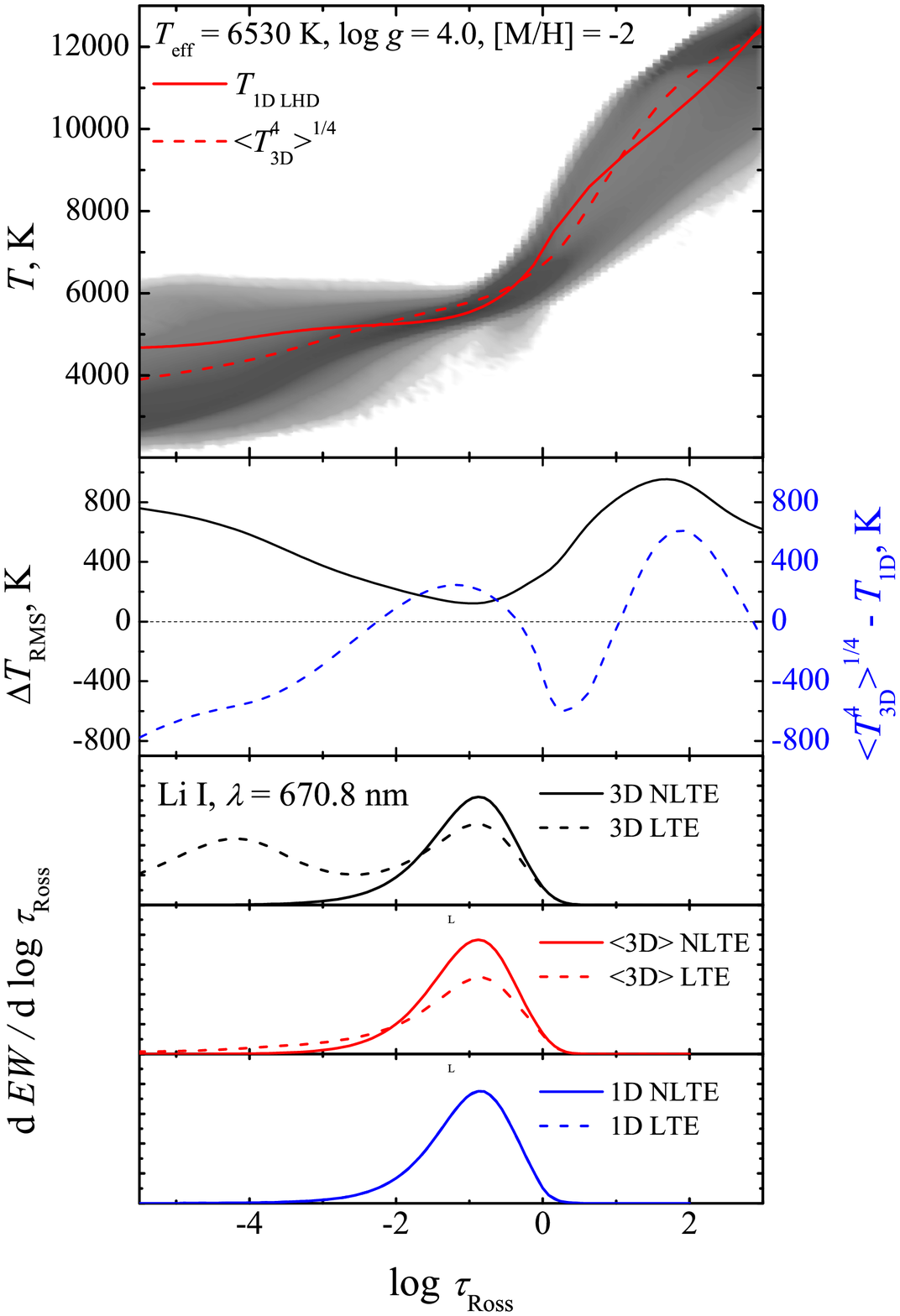}}
\caption{\footnotesize
Same as Fig. \ref{fig:struct-g25}, but for TO stars.}
\label{fig:struct-g40}
\end{figure*}

The results provided in Table~\ref{tab:abucor} and Fig. \ref{fig:abucor} show that, in general, the full $\Delta_{{\rm 3D\,NLTE}~-~{\rm 1D\,LTE}}$ correction (Col.\,5) is different from the $\Delta_{\rm 3D~+\,NLTE}$ ($\equiv \Delta_{{\rm 1D\,NLTE}~-~{\rm 1D\,LTE}} + \Delta_{{\rm 3D\,LTE}~-~{\rm 1D\,LTE}}$) correction (Col.\,4) at all metallicities and in all model atmospheres studied here. We note, however, that while these differences are generally small at solar metallicity (MS\,\#2 is a clear exception), they do indeed become significant at $\moh=-2.0$. In the latter case, the full 3D\,NLTE correction is small and, depending on the line strength, amounts to $\Delta_{{\rm 3D\,NLTE}~-~{\rm 1D\,LTE}} = -0.06 \dots +0.12$~dex. At the same time, the $\Delta_{\rm 3D~+\,NLTE}$ correction is always large and negative, and may reach, for example, to $\sim -0.18\dots-0.23$~dex and $\sim -0.31\dots-0.50$~dex for RGB and TO stars, respectively. On the other hand, the $\Delta_{{\rm \langle3D\rangle\,NLTE - 1D\,LTE}}$ and the $\Delta_{{\rm 1D\,NLTE - 1D\,LTE}}$ abundance corrections are in many cases very similar to the full $\Delta_{{\rm 3D\,NLTE}~-~{\rm 1D\,LTE}}$ correction: the differences are $|\Delta_{{\rm 3D\,NLTE}~-~{\rm 1D\,LTE}} - \Delta_{{\rm \langle3D\rangle\,NLTE - 1D\,LTE}}| \leq 0.04$ and $|\Delta_{{\rm 3D\,NLTE}~-~{\rm 1D\,LTE}} - \Delta_{{\rm 1D\,NLTE - 1D\,LTE}}| \leq 0.08$. Although these discrepancies are small, one should note that full $\Delta_{{\rm 3D\,NLTE}~-~{\rm 1D\,LTE}}$ corrections are small too, especially at $\moh=-2.0$. As a consequence, at lower metallicities the differences $\Delta_{{\rm 3D\,NLTE}~-~{\rm 1D\,LTE}} - \Delta_{{\rm \langle3D\rangle\,NLTE - 1D\,LTE}}$ and $\Delta_{{\rm 3D\,NLTE}~-~{\rm 1D\,LTE}} - \Delta_{{\rm 1D\,NLTE - 1D\,LTE}}$ may become comparable to (or even exceed) the size of the full $\Delta_{{\rm 3D\,NLTE}~-~{\rm 1D\,LTE}}$ corrections.

The behavior of different abundance corrections shown in Table~\ref{tab:abucor} is relatively easy to understand. As seen from Fig.\,\ref{fig:struct-g25} and \ref{fig:struct-g40}, the metal content plays a significant role in shaping the structure of the photospheric layers where the \ion{Li}{i} line forms in RGB and TO stars. At solar metallicity, the mean temperature of the 3D models of RGB and TO stars is close to the 1D radiative equilibrium solution, and the amplitude of the horizontal temperature fluctuations in both cases is moderate, $\Delta T_{\rm RMS} \la 350$~K ($\Delta T_{\rm RMS} = \sqrt{\langle(T - T_0)^2\rangle_{x,y,t}}$, where the angled brackets indicate temporal and horizontal averaging on surfaces of equal optical depth, and $T_0=\langle T \rangle_{x,y,t}$ is the depth-dependent average temperature). In the metal-poor TO model atmosphere, the mean temperature of the 3D model is significantly lower than that predicted in 1D (up to $700$~K), and at the same time the horizontal temperature fluctuations are substantial, increasing with height to $\Delta T_{\rm RMS} \ga 800$~K (Fig.~\ref{fig:struct-g40}, second panels from top). Assuming LTE, both effects lead to significant line strengthening in 3D with respect to 1D in the metal-poor case. This happens because line opacity is a very sensitive non-linear function of temperature. Therefore, even if temperature had a symmetric distribution around the mean value at a given optical depth, the contribution from low-$T$ regions (e.g., inter-granular lanes) towards the total opacity would be more important than that from high-$T$ regions (e.g., granules). As a consequence, the net increase of line opacity in the former would outweigh the net decrease of line opacity in the latter. This would make lines appear stronger in 3D than in $\xtmean{\mbox{3D}}$ or 1D and would result in negative $\Delta_{{\rm 3D\,LTE~-~\langle3D\rangle\,LTE}}$ and $\Delta_{{\rm 3D\,LTE~-~1D\,LTE}}$ abundance corrections \citep[for details see, e.g., ][]{SH02, KSL13}. The situation is similar in the atmosphere of a RGB star, although the line strengthening in 3D is slightly less in this case, because the difference between the temperature profiles of the average $\xtmean{\mbox{3D}}$ and 1D model atmospheres is small, and only the temperature fluctuations contribute to the line strengthening. Qualitatively, the effect is similar in all other model atmospheres studied here. Note, however, that the situation is different in NLTE (see below).

Comparison of the line contribution functions obtained in LTE (see Fig.~\ref{fig:struct-g25} and \ref{fig:struct-g40}) reveals their close similarity at $\moh=0.0$. Consequently, one may infer that the effect of the horizontal temperature fluctuations is small at solar metallicity. However, the line contribution functions in 3D\,LTE and $\xtmean{\mbox{3D}}$\,LTE are significantly different at $\moh=-2.0$, both in RGB and TO stars, which shows that the role of horizontal fluctuations now becomes substantial and dominates over the effect of  the reduced mean temperature. In the metal-poor 3D hydrodynamical models, the fluctuations produce regions in the atmosphere where temperature drops significantly below that predicted by the 1D model. This leads to larger concentration of \ion{Li}{i} in the 3D models, and thus, to stronger lines in 3D\,LTE (i.e., with respect to those in 1D\,LTE) and negative $\Delta_{{\rm 3D\,LTE}~-~{\rm 1D\,LTE}}$ abundance corrections. In NLTE (both 1D and 3D), however, atomic level population numbers are more sensitive to the average radiation field than to local temperature \citep[e.g.,][]{CS2000,ACB03}. As a consequence, \ion{Li}{i} gets significantly over-ionized  in the outer atmosphere. Therefore, the mean concentration of \ion{Li}{i} with respect to what would be expected in LTE is reduced. This leads to weaker spectral lines in 3D\,NLTE than in 3D\,LTE, and thus, significantly reduced $\Delta_{{\rm 3D\,NLTE}~-~{\rm 1D\,LTE}}$ abundance corrections which are similar in magnitude to $\Delta_{{\rm 1D\,NLTE}~-~{\rm 1D\,LTE}}$ corrections.

It is important to note that, in general, the choice of the microtubulence velocity used in the spectral line synthesis with the average $\xtmean{\mbox{3D}}$ and 1D model atmospheres may influence the corresponding abundance corrections. Fortunately, the resonance line of lithium  observed in stars covered by our atmospheric parameter range is relatively weak, and so the influence of microturbulence on the line strength is, in fact, minor. Our tests show that the difference in the abundance correction obtained at microturbulent velocities of 1.0~km/s and 5.0~km/s is small, $\lesssim0.05$~dex. The difference in abundance corrections computed at more typical values of 1.0 and 2.0~km/s was less than 0.02~dex, irrespective of whether the line formation was treated in NLTE or LTE, even for lines as strong as $W \approx 8$\,pm. This illustrates that saturation of the stronger line of the \ion{Li}{i} 670.8 nm doublet is not significant in the range of $W$s investigated. However, in the Li-rich stars, this line may become saturated and thus experience  an enhanced sensitivity to microturbulence and larger non-LTE effects due to photon losses \citep[e.g.,][]{LAB09}.

\begin{figure}[tb]
\centering
\mbox{\includegraphics[width=8.5cm]{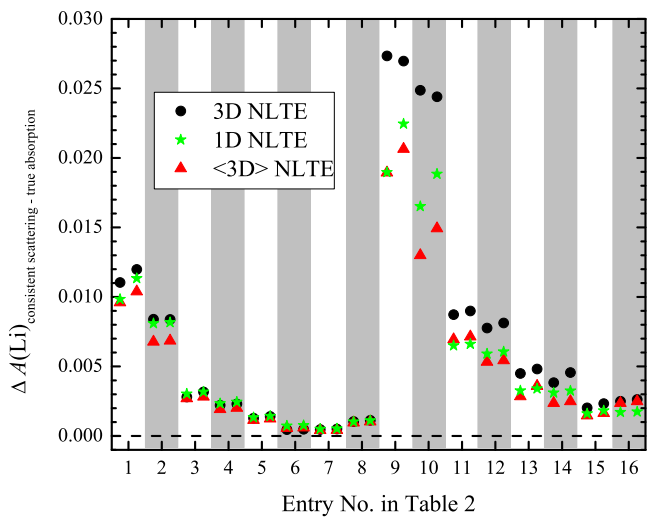}}
\caption{
Changes in $A$(Li) when different treatments of continuum scattering are applied in calculations of the departure coefficients (coherent scattering versus true absorption). Differently shaded areas represent different model atmospheres as they are listed in Table \ref{tab:abucor}. Symbols on the left side of each shaded region represent spectral lines with $W = 0.5$ pm, while symbols on the right represent those with $W = 8.0$ pm.}
\label{fig:scattering}
\end{figure}

To check the importance of continuum scattering for the \ion{Li}{i} statistical equilibrium solution, we  computed an additional set of synthetic spectra where the departure coefficients were computed by treating scattering as true absorption. The effect of this change on the estimates of Li abundance in 3D, $\xtmean{\mbox{3D}}$, and 1D is relatively small\footnote{Convergence limits for the computations of $J_\nu$ and $W$ were set at 0.01~\% and 0.001~\%, respectively, to ensure the precision of 0.001~dex or better in the computed abundance corrections.}, as shown in Fig.\,\ref{fig:scattering}: at solar metallicity, the differences in abundance increase from about $0.001$ dex in the dwarfs to up to $0.012$~dex in the giants, while a more pronounced effect of scattering is seen in metal-poor stars, where the abundance differences range from $0.005$ dex in the dwarfs to $0.028$~dex in the case of the metal-poor red giants. This is because, especially in the metal-poor giants, the opacity coefficients of continuum scattering and true absorption become comparable in magnitude. However, the effect on the abundance corrections (e.g., $\Delta_{{\rm 3D\,NLTE}~-~{\rm 1D\,NLTE}}$) is much smaller and does not exceed $0.002$~dex, except for the two metal-poor giants. Qualitatively, these findings are similar to those obtained by \citet[][]{HAC11} who found that the impact of scattering on the formation of fictitious \ion{Fe}{i} resonance lines in the atmosphere of red giant star is negligible at 500~nm ($\moh=-2.0$ and 0.0), both in 3D and 1D, with the resulting differences in the abundance corrections of $<0.015$~dex. In general, we find that 3D spectral line formation is slightly more sensitive to scattering effects compared to 1D, such that the abundance corrections are slightly more positive when continuum scattering is treated consistently.

General uncertainties involved in 3D\,NLTE spectral synthesis (excluding uncertainties in atomic data) can be illustrated by comparing Li abundance corrections, $\Delta_{{\rm 3D\,NLTE}~-~{\rm 1D\,LTE}}$, computed in this work with those obtained using the interpolation formula from \citet[][Eq.~B.1]{SBC10}. This formula is based on the 3D~NLTE and 1D~LTE spectral synthesis computations obtained using a grid of 3D hydrodynamical \COBOLD\ and 1D hydrostatic \LHD\ model atmospheres of dwarfs and subgiant stars ($\teff\approx5470\dots6560$~K, $\log g=3.5\dots4.5$, $\moh=-2.0$). The spectral synthesis computations of \citet[][]{SBC10} utilize departure coefficients computed with a simpler Li model atom and an older version of the \NLTE\ code, treating continuum scattering as true absorption. Their formula is supposed to have an internal precision of $\approx \pm0.005$~dex as long as extrapolation towards the values of equivalent widths and atmospheric parameters that are not covered in their simulations is avoided \citep[see Fig.~6 and 7 in][]{SBC10}. For this comparison we used models \#12, 13, 14, and 15 from our grid, because their atmospheric parameters fall in the range covered by the simulations of \citet{SBC10}. We find that corrections obtained in this work differ from those computed using the formula of \citet{SBC10} by $-0.027\pm0.012$~dex on average for lines with $W$ = $5$~pm and by $-0.055\pm0.017$~dex for lines with $W$ = $8$~pm, without a notable systematic dependence on atmospheric parameters. The larger disagreement in the case of stronger lines is not unexpected since, in the present calculations of the departure coefficients, we have taken the doublet fine structure of the 670.8~nm line into account, while the single component approximation used by \citet{SBC10} is only valid for weak lines, and consequently \citet{SBC10} does not recommend using their interpolation formula when saturation effects become important.

\section{Summary and conclusions\label{sect:conclus}}

We studied whether various simplified approaches (such as deriving abundances in 1D\,NLTE and applying to them 3D--1D\,LTE abundance corrections; or using $\xtmean{\mbox{3D}}$ models with the 1D\,NLTE analysis tools) could be used to substitute for the full 3D\,NLTE methodology to  derive  lithium abundances in the atmospheres of RGB, SGB, TO, and MS stars. Our results clearly show that the application of the combined $\Delta_{\rm 3D~+\,NLTE}$ correction may lead to erroneous results in the case of lithium at sub-solar metallicities with all model atmospheres studied here. On the other hand, both $\xtmean{\mbox{3D}}$\,NLTE and 1D\,NLTE approaches provide a good approximation to the full 3D\,NLTE solution.

Our results suggest that, in most cases, $\xtmean{\mbox{3D}}$\,NLTE or 1D\,NLTE modeling should be sufficient for deriving lithium abundances accurate to $\sim0.1$~dex, i.e., with respect to those obtained in 3D\,NLTE. Nevertheless, since the full $\Delta_{{\rm 3D\,NLTE}~-~{\rm 1D\,LTE}}$ correction, in certain cases, is only larger than 0.1~dex, application of $\Delta_{{\rm 1D\,NLTE - 1D\,LTE}}$ or $\Delta_{{\rm \langle3D\rangle\,NLTE - 1D\,LTE}}$ corrections may lead to systematical bias that is comparable to the size of the $\Delta_{{\rm 3D\,NLTE}~-~{\rm 1D\,LTE}}$ correction itself. This may be obviously unacceptable when high accuracy/reliability in lithium abundances is needed. Ideally, the full 3D\,NLTE corrections should be preferred over any of the approximate approaches.

The use of $\xtmean{\mbox{3D}}$\,NLTE or 1D\,NLTE approaches, moreover, cannot be extended in a straightforward fashion to other chemical elements: the importance of different physical processes leading to the departures from LTE may vary from one chemical element to another. As we have seen from the results obtained in our study (which are in line with the earlier findings of \citealt{ACB03}), the 3D--1D and NLTE--LTE abundance corrections generally tend to compensate each other, so that the total $\Delta_{{\rm 3D\,NLTE}~-~{\rm 1D\,LTE}}$ abundance correction is small. This, however, seems not to be the case with other elements where this kind of compensation does not occur. For example, in their study of NLTE \ion{Fe}{i} spectral line formation based on a comparison with the observed spectra, \citet[][]{MLK13} find the $\xtmean{\mbox{3D}}$\,NLTE approach is only applicable to a limited subset of \ion{Fe}{i} lines. Therefore, verification with the full 3D\,NLTE approach is needed on a case-by-case basis before any simplified approach may be applied with confidence to the abundance analysis of other elements in the atmospheres of real stars.

\begin{acknowledgements}

We thank K.~Lind for valuable comments and suggestions that helped to improve the paper. We are grateful to P.~Bonifacio, V.~Dobrovolskas, and D.~Prakapavi\v{c}ius for their help during various stages of the paper's preparation. This work was supported by grants from the Research Council of Lithuania (MIP-065/2013) and the bilateral French-Lithuanian program Gilibert (TAP~LZ~06/2013, Research Council of Lithuania; 28471NE, Campus France). HGL and MS acknowledge funding from the Research Council of Lithuania for  research visits to Vilnius. HGL acknowledges financial support by the Sonderforschungsbereich SFB 881 "The Milky Way System" (subprojects A4, A5) of the German Research Foundation (DFG). EC is grateful to the FONDATION MERAC for funding her fellowship. Some model computations were performed using resources at the High Performance Computing Center, HPC Sauletekis, of the Faculty of Physics, Vilnius University.

\end{acknowledgements}

\bibliographystyle{aa}

\begin{appendix}

\section{Model atom of lithium\label{app:atom}}

The data we used to update the model atom of \ion{Li}{i} are provided in Tables~\ref{tab:Liatom_levels} and \ref{tab:Liatom_transitions}. We list atomic parameters of the energy levels and/or transitions, as well as the sources from which the data were taken.

\begin{table}[b]

\caption{Energy levels of the \ion{Li}{i} model atom. All data are from the
NIST database.
\label{tab:Liatom_levels}}
\centering
\begin{tabular}{cccc}

\hline\noalign{\smallskip}
Level & Configuration & Energy & Statistical \\
  \#  &               &  [eV]  & weight      \\
\noalign{\smallskip}
\hline
\noalign{\smallskip}

  1   &      2s       & 0.000  &     2       \\
  2   &      2p       & 1.848  &     6       \\
  3   &      3s       & 3.373  &     2       \\
  4   &      3p       & 3.834  &     6       \\
  5   &      3d       & 3.879  &     10      \\
  6   &      4s       & 4.341  &     2       \\
  7   &      4p       & 4.522  &     6       \\
  8   &      4d       & 4.541  &     10      \\
  9   &      4f       & 4.542  &     14      \\
  10  &      5s       & 4.749  &     2       \\
  11  &      5p       & 4.837  &     6       \\
  12  &      5d       & 4.847  &     10      \\
  13  &      5f       & 4.848  &     14      \\
  14  &      6s       & 4.958  &     2       \\
  15  &      6p       & 5.008  &     6       \\
  16  &      6d       & 5.014  &     10      \\
  17  &      6f       & 5.014  &     14      \\
  18  &      7s       & 5.079  &     2      \\
  19  &      7p       & 5.110  &     6      \\
  20  &      7d       & 5.114  &     10      \\
  21  &      8s       & 5.156  &     2      \\
  22  &      8p       & 5.177  &     6      \\
  23  &      8d       & 5.179  &     10      \\
  24  &      9s       & 5.208  &     2      \\
  25  &      9p       & 5.222  &     6      \\
  26  &      9d       & 5.224  &     10      \\

\noalign{\smallskip}
\hline
\noalign{\smallskip}

\end{tabular}
\end{table}

\onecolumn
\centering
\tablehead{
\hline\noalign{\smallskip}
Transition & Lower & Upper & Transition    & $\lambda$    & \multicolumn{3}{c}{Transition probability} \\
\#         & level & level & configuration & vacuum [\AA] & $A_{j\,i}$ & $B_{i\,j}$ & $f_{i\,j}$ \\
\noalign{\smallskip}
\hline
\noalign{\smallskip}}
\tabletail{
\noalign{\smallskip}
\hline
\multicolumn{8}{c}{\small\sl continued on next page}\\
\noalign{\smallskip}}
\tablelasttail{\hline}
\tablecaption[]{Radiative bound-bound transitions of the \ion{Li}{i} model atom. Data is taken from the NIST database.}
\begin{supertabular}{cccccccc}
\label{tab:Liatom_transitions}
1 & 1 & 2 & 2s-2p & 6709.7       & 3.69E+07 & 8.41E+10 & 7.47E-01 \\
2 & 1 & 4 & 2s-3p & 3233.6       & 1.00E+06 & 2.56E+08 & 4.71E-03 \\
3 & 1 & 7 & 2s-4p & 2742.0       & 1.25E+06 & 1.94E+08 & 4.22E-03 \\
4 & 2 & 3 & 2p-3s & 8128.6       & 3.35E+07 & 1.51E+10 & 1.10E-01 \\
5 & 2 & 5 & 2p-3d & 6105.3       & 6.86E+07 & 6.55E+10 & 6.39E-01 \\
6 & 2 & 6 & 2p-4s & 4973.1       & 1.04E+07 & 1.07E+09 & 1.28E-02 \\
7 & 2 & 8 & 2p-4d & 4604.1       & 2.32E+07 & 9.51E+09 & 1.23E-01 \\
8 & 3 & 4 & 3s-3p & 26887               & 3.74E+06 & 5.48E+11 & 1.22E+00 \\
9 & 3 & 7 & 3s-4p & 10795               & 6.80E+02 & 6.46E+06 & 3.56E-05 \\
10 & 4 & 8 & 3p-4d & 17550       & 6.78E+06 & 1.54E+11 & 5.22E-01 \\
11 & 5 & 7 & 3d-4p & 19281       & 5.36E+05 & 5.81E+09 & 1.79E-02 \\
12 & 1 & 11 & 2s-5p & 2563.1    & 8.87E+05 & 1.13E+08 & 2.62E-03 \\
13 & 1 & 15 & 2s-6p & 2475.8    & 5.73E+05 & 6.57E+07 & 1.58E-03 \\
14 & 2 & 12 & 2p-5d & 4133.7    & 1.09E+07 & 3.23E+09 & 4.65E-02 \\
15 & 2 & 14 & 2p-6s & 3986.6    & 2.59E+06 & 1.38E+08 & 2.06E-03 \\
16 & 2 & 10 & 2p-5s & 4274.3    & 4.75E+06 & 3.11E+08 & 4.34E-03 \\
17 & 4 & 6 & 3p-4s & 24469       & 7.45E+06 & 9.16E+10 & 2.23E-01 \\
18 & 4 & 10 & 3p-5s & 13560      & 2.83E+06 & 5.92E+09 & 2.60E-02 \\
19 & 4 & 12 & 3p-5d & 12240      & 3.48E+06 & 2.68E+10 & 1.30E-01 \\
20 & 4 & 14 & 3p-6s & 11034      & 1.46E+06 & 1.65E+09 & 8.88E-03 \\
21 & 4 & 16 & 3d-5p & 10513      & 1.97E+06 & 9.60E+09 & 5.44E-02 \\
22 & 5 & 9 & 3d-4f & 18701       & 1.38E+07 & 3.18E+11 & 1.02E+00 \\
23 & 5 & 11 & 3d-5p & 12932      & 2.28E+05 & 7.45E+08 & 3.43E-03 \\
24 & 5 & 13 & 3d-5f & 12785      & 4.58E+06 & 3.37E+10 & 1.57E-01 \\
25 & 6 & 7 & 4s-4p & 68611       & 7.76E+05 & 1.89E+12 & 1.64E+00 \\
26 & 6 & 11 & 4s-5p & 24978      & 3.38E+03 & 3.98E+08 & 9.48E-04 \\
27 & 6 & 15 & 4s-6p & 18591      & 1.87E+03 & 9.07E+07 & 2.91E-04 \\
28 & 7 & 10 & 4p-5s & 54646      & 2.25E+06 & 3.08E+11 & 3.36E-01 \\
29 & 7 & 12 & 4p-5d & 38089      & 1.36E+06 & 3.15E+11 & 4.93E-01 \\
30 & 7 & 14 & 4p-6s & 28422      & 9.59E+05 & 1.85E+10 & 3.87E-02 \\
31 & 7 & 16 & 4p-6d & 25203      & 8.38E+05 & 5.63E+10 & 1.33E-01 \\
32 & 8 & 11 & 4d-5p & 41802      & 2.76E+05 & 3.04E+10 & 4.34E-02 \\
33 & 8 & 13 & 4d-5f & 40305      & 2.58E+06 & 5.95E+11 & 8.80E-01 \\
34 & 8 & 15 & 4d-6p & 26543      & 1.37E+05 & 3.87E+09 & 8.68E-03 \\
 35 &  1  &  19 & 2s-7p & 2426.2     & 3.82E+05 & 4.12E+07 & 1.01E-03 \\
 36 &  1  &  22 & 2s-8p & 2395.1     & 2.66E+05 & 2.76E+07 & 6.87E-04 \\
 37 &  1  &  25 & 2s-9p & 2374.3     & 1.92E+05 & 1.94E+07 & 4.86E-04 \\
 38 &  2  &  14 & 2p-6d & 3916.4     & 5.96E+06 & 3.17E+08 & 4.73E-03 \\
 39 &  2  &  18 & 2p-7s & 3836.7     & 1.56E+06 & 7.39E+07 & 1.15E-03 \\
 40 &  2  &  20 & 2p-7d & 3796.1     & 3.65E+06 & 8.37E+08 & 1.31E-02 \\
 41 &  2  &  21 & 2p-8s & 3747.7     & 1.01E+06 & 4.46E+07 & 7.09E-04 \\
 42 &  2  &  23 & 2p-8d & 3722       & 2.41E+06 & 3.14E+08 & 5.02E-03 \\
 43 &  2  &  26 & 2p-9d & 3672.8     & 1.68E+06 & 3.49E+08 & 5.66E-03 \\
 44 &  3  &  11 & 3s-5p & 8467.8     & 4.04E+04 & 1.85E+08 & 1.30E-03 \\
 45 &  3  &  15 & 3s-6p & 7584.5     & 4.38E+04 & 1.44E+08 & 1.13E-03 \\
 46 &  3  &  19 & 3s-7p & 7137.1     & 3.60E+04 & 9.88E+07 & 8.25E-04 \\
 47 &  3  &  22 & 3s-8p & 6875       & 2.79E+04 & 6.85E+07 & 5.93E-04 \\
 48 &  4  &  18 & 3p-7s & 9958       & 8.62E+05 & 7.14E+08 & 4.27E-03 \\
 49 &  4  &  20 & 3p-7d & 9689       & 1.23E+06 & 4.69E+09 & 2.89E-02 \\
 50 &  4  &  21 & 3p-8s & 9379       & 5.54E+05 & 3.84E+08 & 2.44E-03 \\
 51 &  4  &  23 & 3p-8d & 9220       & 8.09E+05 & 2.66E+09 & 1.72E-02 \\
 52 &  5  &  15 & 3d-6p & 10980      & 1.19E+05 & 2.38E+08 & 1.29E-03 \\
 53 &  5  &  19 & 3d-7p & 10066      & 7.08E+04 & 1.09E+08 & 6.45E-04 \\
 54 &  5  &  22 & 3d-8p & 9552.4     & 4.57E+04 & 6.02E+07 & 3.75E-04 \\
 55 &  6  &  19 & 4s-7p & 16115      & 4.15E+03 & 1.31E+08 & 4.85E-04 \\
 56 &  6  &  22 & 4s-8p & 14838      & 4.40E+03 & 1.09E+08 & 4.36E-04 \\
 57 &  7  &  18 & 4p-7s & 22230      & 5.37E+05 & 4.95E+09 & 1.33E-02 \\
 58 &  7  &  20 & 4p-7d & 20934      & 5.31E+05 & 2.04E+10 & 5.81E-02 \\
 59 &  7  &  21 & 4p-8s & 19541      & 3.36E+05 & 2.10E+09 & 6.41E-03 \\
 60 &  7  &  23 & 4p-8d & 18862      & 3.55E+05 & 9.99E+09 & 3.16E-02 \\
 61 &  8  &   9 & 4d-4f & 1.47E+07   & 6.90E-02 & 7.68E+11 & 3.12E-03 \\
 62 &  8  &  19 & 4d-7p & 21768      & 7.80E+04 & 1.22E+09 & 3.32E-03 \\
 63 &  8  &  22 & 4d-8p & 19500      & 4.91E+04 & 5.50E+08 & 1.68E-03 \\
 64 &  9  &  16 & 4f-6d & 26267      & 2.50E+04 & 8.15E+08 & 1.85E-03 \\
 65 & 10  &  11 & 5s-5p & 1.3965E+05 & 2.34E+05 & 4.81E+12 & 2.05E+00 \\
 66 & 10  &  15 & 5s-6p & 47816      & 3.33E+03 & 2.75E+09 & 3.42E-03 \\
 67 & 10  &  22 & 5s-8p & 28968      & 4.63E+02 & 8.50E+07 & 1.75E-04 \\
 68 & 11  &  12 & 5p-5d & 1.26E+06   & 4.80E+02 & 4.03E+12 & 1.90E-01 \\
 69 & 11  &  14 & 5p-6s & 1.0287E+05 & 8.49E+05 & 7.75E+11 & 4.49E-01 \\
 70 & 11  &  16 & 5p-6d & 70335      & 3.99E+05 & 5.82E+11 & 4.93E-01 \\
 71 & 11  &  18 & 5p-7s & 51221      & 3.89E+05 & 4.39E+10 & 5.10E-02 \\
 72 & 11  &  20 & 5p-7d & 44824      & 2.72E+05 & 1.03E+11 & 1.37E-01 \\
 73 & 11  &  21 & 5p-8s & 38887      & 2.32E+05 & 1.14E+10 & 1.75E-02 \\
 74 & 11  &  23 & 5p-8d & 36288      & 1.86E+05 & 3.73E+10 & 6.12E-02 \\
 75 & 12  &  13 & 5d-5f & 1.05E+07   & 6.96E-01 & 2.85E+12 & 1.61E-02 \\
 76 & 12  &  15 & 5d-6p & 77166      & 1.37E+05 & 9.51E+10 & 7.34E-02 \\
 77 & 12  &  19 & 5d-7p & 47116      & 7.47E+04 & 1.18E+10 & 1.49E-02 \\
 78 & 12  &  22 & 5d-8p & 37641      & 4.54E+04 & 3.66E+09 & 5.79E-03 \\
 79 & 13  &  16 & 5f-6d & 75029      & 4.19E+04 & 3.18E+10 & 2.53E-02 \\
 80 & 14  &  15 & 6s-6p & 2.4802E+05 & 8.88E+04 & 1.02E+13 & 2.46E+00 \\
 81 & 14  &  19 & 6s-7p & 81320      & 2.17E+03 & 8.82E+09 & 6.46E-03 \\
 82 & 15  &  16 & 6p-6d & 2.15E+06   & 2.11E+02 & 8.81E+12 & 2.44E-01 \\
 83 & 15  &  18 & 6p-7s & 1.7329E+05 & 3.74E+05 & 1.63E+12 & 5.61E-01 \\
 84 & 15  &  20 & 6p-7d & 1.1687E+05 & 1.48E+05 & 9.91E+11 & 5.05E-01 \\
 85 & 15  &  21 & 6p-8s & 83594      & 1.81E+05 & 8.87E+10 & 6.32E-02 \\
 86 & 15  &  23 & 6p-8d & 72440      & 1.07E+05 & 1.71E+11 & 1.40E-01 \\
 87 & 16  &  19 & 6d-7p & 1.282E+05  & 7.08E+04 & 2.25E+11 & 1.05E-01 \\
 88 & 16  &  22 & 6d-8p & 76086      & 4.16E+04 & 2.77E+10 & 2.17E-02 \\
 89 & 18  &  19 & 7s-7p & 4.0088E+05 & 3.96E+04 & 1.93E+13 & 2.86E+00 \\
 90 & 18  &  22 & 7s-8p & 1.2760E+05 & 1.34E+03 & 2.10E+10 & 9.81E-03 \\
 91 & 19  &  20 & 7p-7d & 3.43E+06   & 9.97E+01 & 1.69E+13 & 2.93E-01 \\
 92 & 19  &  21 & 7p-8s & 2.7045E+05 & 1.84E+05 & 3.05E+12 & 6.73E-01 \\
 93 & 19  &  23 & 7p-8d & 1.8052E+05 & 6.38E+04 & 1.57E+12 & 5.19E-01 \\
 94 & 20  &  22 & 7d-8p & 1.9797E+05 & 3.93E+04 & 4.60E+11 & 1.39E-01 \\
 95 & 21  &  22 & 8s-8p & 6.0779E+05 & 1.96E+04 & 3.32E+13 & 3.26E+00 \\
 96 & 22  &  23 & 8p-8d & 5.08E+06   & 5.32E+01 & 2.94E+13 & 3.44E-01 \\
\end{supertabular}
\twocolumn

\end{appendix}

\end{document}